\documentclass[twocolumn,prl,aps,showpacs]{revtex4}
\usepackage{epsfig}
\usepackage{fancyhdr}

\newcommand{\beq}{\begin{equation}}
\newcommand{\eeq}{\end{equation}}
\begin{document}

\title{Disorder in DNA-Linked Gold Nanoparticle Assemblies}

\author{Nolan C.\ Harris and Ching-Hwa Kiang$^*$}

\affiliation{Department of Physics and Astronomy,
Rice University, Houston, TX\ \ 77005--1892}

\begin{abstract}
We report experimental observations on the effect of disorder on the
phase behavior of DNA-linked nanoparticle assemblies.  Variation in
DNA linker lengths results in different melting temperatures of the
DNA-linked nanoparticle assemblies. We observed an unusual trend of a
non-monotonic ``zigzag'' pattern in the melting temperature as a function of DNA
linker length.  Linker DNA resulting in unequal DNA duplex lengths
introduces disorder and lowers the melting temperature of the nanoparticle system.
Comparison with free DNA thermodynamics shows that such an
anomalous zigzag pattern does not exist for free DNA duplex melting,
which suggests that the disorder introduced
by unequal DNA duplex lengths results in this unusual collective
behavior of DNA-linked nanoparticle assemblies.
\end{abstract}



\maketitle
\thispagestyle{fancy} \chead{PHYSICAL REVIEW LETTERS} \lfoot{\thepage} \cfoot{} \rfoot{\thepage}

DNA-linked nanoparticle assemblies are a novel system in which
gold nanoparticles are chemically affixed to known DNA sequences
to create DNA ``probes'' with the capability to self-assemble into
aggregates~\cite{Mirkin03,Kiang03a,Kiang05a,Kiang05b,Kiang05d}.  The interaction potential
between these colloids are tunable and controllable, which makes these
multicomponent complex fluids particularly suitable for
studying the link between the interaction potential and
phase behavior ~\cite{Frenkel04a,Russel03}.
Similar colloidal phase transitions have also been used to
detect the protein interactions at membrane surfaces \cite{Groves04a}.
On the other hand, the dynamics of DNA melting and hybridization
in DNA replication and transcription is also a subject of
intense investigation \cite{Libchaber03a,Zocchi03a}.

The change in optical property upon aggregation makes DNA-linked
nanoparticle systems a potential tool in future DNA detection
technology~\cite{Mirkin98,Mirkin00b}. DNA detection is important
in medical research for applications such as detection of genetic
diseases, RNA profiling, and biodefense
~\cite{Whitten03a,Lockhart00,Hill00a,Janda02a,Lu04a}.  The DNA-linked
nanoparticle detection systems utilize the sequence-dependent hybridization
of DNA for accuracy and the optical properties of colloidal
gold for sensitivity to create a DNA detection method that
changes color upon the introduction of a specific DNA sequence.
Study of these DNA-gold nanoparticle assemblies is warranted to gain a
fundamental understanding of DNA hybridization in confined
geometries, as well as to probe the potential of this and similar
systems for practical applications in biotechnology
~\cite{Boal00,Kramer03}.

Much like other colloidal suspensions
~\cite{Weitz04a,Weitz04b,Cipelletti04}, the DNA-gold nanoparticle system
has been shown to exhibit interesting phase behavior. The assembly
melting temperature is dependent upon parameters such as particle
size, DNA composition, and electrolyte
concentrations~\cite{Kiang03a,Mirkin03}. Recent theory has sought
to explain these observations
~\cite{Frenkel04a,Stroud03b,Tkachenko02}.  More study
is necessary to fully understand the underlying mechanisms affecting
the phase behavior of this system.

\begin{figure}[!b]
\begin{center}
\vspace{-0.1in}
\epsfig{file=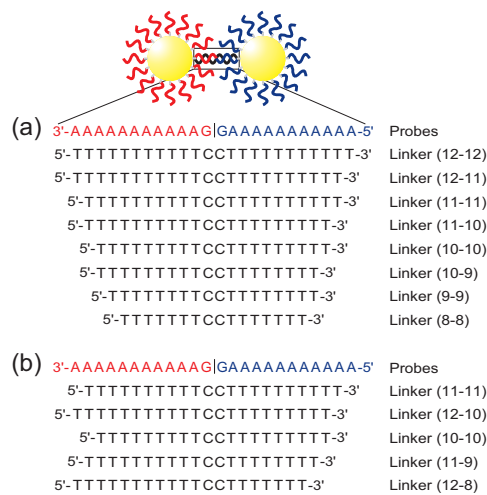,clip=,width=2.5in}
\vspace{-0.2in}
\end{center}
\caption{(color online) Basic building block of DNA-linked gold nanoparticle
aggregates.  There are no bonds between the 2 G bases in
the probe DNA.  Linker DNA sequences used are the following: (a) sequences of
varying length and (b) sequences with the same length, but
different base distribution. All linkers are named such that the
number of bases from the 5$^\prime$ end to the middle of the
sequence are listed first, followed by the number of bases from
the 3$^\prime$ end to the middle.}
\label{fig:sequences}
\end{figure}

Here we report experimental observations of the effects of
disorder on the melting temperature of the DNA-linked nanoparticle
assemblies \cite{Kiang03a,Kiang05a}.
The basic building block is illustrated in Fig.~\ref{fig:sequences}.
Disorder in the DNA duplex length was introduced by
choosing linker DNA with an odd number of bases.
We studied the melting temperature as a
function of DNA linker length.  The melting temperature
accesses the stability of systems.
DNA gold nanoparticle probes were synthesized using previously
described methods~\cite{Kiang03a}.
DNA-gold aggregates were formed by mixing probe DNA particles with
linker DNA~\cite{Kiang03a,Kiang05a,Kiang05b}.  Upon the addition of linker DNA,
the solutions were allowed to stand at 4 $^\circ$C for several
days in order to achieve maximum network aggregation.

Simple DNA sequences with uniform base composition were chosen
in order to remove complications resulting from
sequence-dependent effects. The probe sequences used in these
experiments consisted of 11 identical bases in a row and one
``discriminator'' base at the terminus that is not thiol
modified (see Fig.~\ref{fig:sequences}). In this system, when
gold-attached DNA probes aligned to bind with a specific linker
sequence, the two ``G'' bases at the end of each probe met in
the middle. These bases served to ensure that every linker
sequence hybridized by aligning with the two G bases in the
center. Hybridizing any other way will result in at least two
base pair mismatches, which is much less stable than a fully
complementary bond and, therefore, will not be present in any
significant amount in our aggregates.

We chose linker DNA sequences to observe the effect of linker
length on the DNA-gold nanoparticle system. Gold-attached DNA
clusters were prepared using linkers that ranged from 24 to 16
bases in length.  The linker concentration was adjusted to be
oversaturated so the melting temperature does not depend on
linker concentration \cite{Frenkel04a,Safran03a}. Typical 10 nm
gold and DNA linker concentrations were $\sim$4$\times$10$^{17}$
particles/l and 7$\times$10$^{-6}$ M, respectively.  Melting
curves of corresponding DNA free of gold nanoparticles were
measured for comparison \cite{Kiang03a,Kiang05a}.  Melting of
the system is observed at 260 nm, heating the solution at a rate
of 1 $^\circ$C/min.

\begin{figure}[!b]
\begin{center}
\epsfig{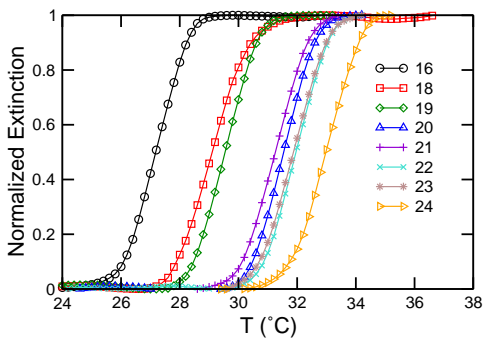}
\vspace{-0.2in}
\end{center}
\caption{(color online) Melting curves at 260nm for
gold-attached DNA assemblies formed using linkers with between
16 and 24 bases.} \label{fig:meltcurve}
\end{figure}

\begin{figure}[!b]
\begin{center}
\epsfig{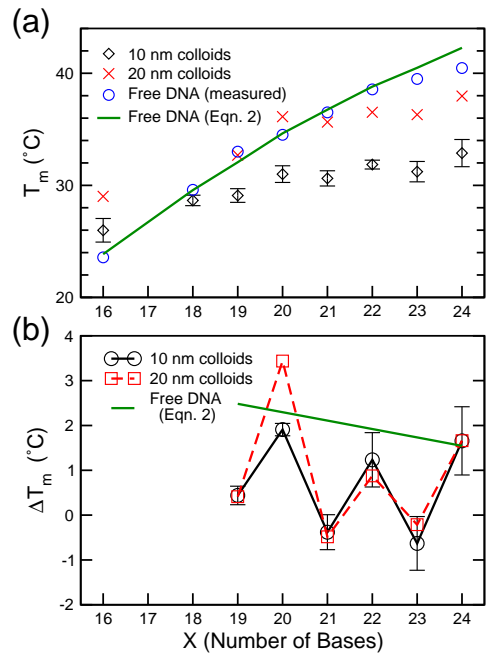}
\vspace{-0.2in}
\end{center}
\caption{(color online) (a) Melting temperature as a function of
linker sequence length for free, unattached DNA, and 10 and 20
nm gold-attached DNA. Solid line represents predicted values
using empirically determined thermodynamic parameters. (b)
Change in melting temperature as linker sequence length
increases. Values and error bars (one standard error) for 10 nm
gold-attached DNA data are taken from four separate
experiments.} \label{fig:Tm}
\end{figure}

Experimental melting curves detailing the typical melting
behavior of DNA-linked nanoparticle networks, in which linkers of
varying length were used, are given in Fig.~\ref{fig:meltcurve}.
>From the data it is evident that the melting of these aggregates
produce a sharp melting transition.  More interestingly, we
noticed that linker sequences with 21 and 23 bases melted at
lower temperatures than linkers with 20 and 22 bases,
respectively, as shown in Fig.~\ref{fig:Tm}(a).

To identify the cause of this anomaly in $T_m$, we compared the
thermodynamic parameters of corresponding DNA that are not
attached to gold nanoparticles. The experimental results of
using free DNA counterparts in similar experimental conditions
are shown in Fig.~\ref{fig:Tm}(a).
The melting temperature of a free DNA duplex increases as the
number of base pairs increases and does not display this anomaly.
In contrast, the melting
temperature of gold-attached DNA duplexes oscillates. $T_m$
increases with linker length when a linker has an even number of
bases and decreases when it has an odd number. This behavior is
also observed on a system with a different particle size, as
demonstrated in Fig.~\ref{fig:meltcurve}, where the $T_m$ data for
aggregates formed using 20 nm gold particles have a similar trend.

The melting temperatures of free DNA duplexes were predicted using
experimental thermodynamic parameters. The thermodynamic relationship
describing the free energy of a system is given by
\beq \Delta G^0 = \Delta H^0 - T \Delta S^0.
\label{2ndlaw}
\eeq
The melting temperature of a DNA duplex is defined as the midpoint
between aggregated and dispersed phases in the absorption curves
\cite{CantorII}. At that temperature the concentration of single-stranded
DNA is equal to that of double-stranded DNA. A widely used equation to
calculate $T_m$, taking into account influence from nearest
neighbor interactions and salt concentration, gives the melting
temperature of a duplex as \cite{Sugimoto96,Breslauer86}

{\small
\beq T_m = \left( {\Delta H^0 +~3.4~{{\mbox {kcal} \over \mbox {mol}}} \over
\Delta S^0 - R~\mbox {ln} \left( {1 \over \mbox {[DNA]}} \right)
} \right) + 16.6 \: \mbox {log}_{10} ([\mbox {Na}^+]).
\label{Tm} \eeq}

The change in enthalpy, $\Delta H^0$, and the change in entropy,
$\Delta S^0$, for a given DNA duplex were calculated by using
nearest neighbor parameters \cite{Sugimoto96,Breslauer86}. The
free energy change,  $\Delta G^0$, was estimated using
Eq.~(\ref{2ndlaw}).  A sample calculation for a DNA duplex
formed using two 12 base probes and 16 base linker DNA is
illustrated below.
\begin{figure}[!h]
\begin{center}
\epsfig{file=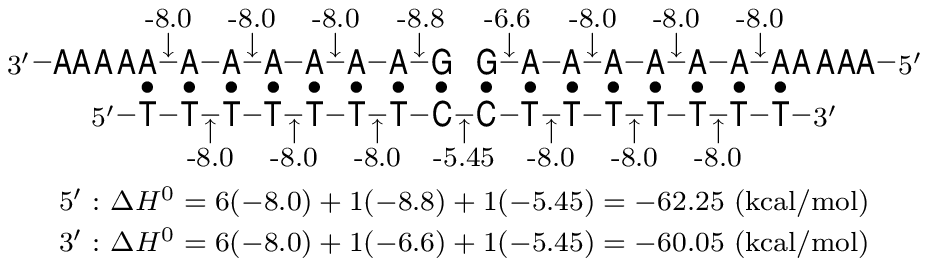,clip=,width=.95\columnwidth}
\vspace{-0.2in}
\end{center}
\end{figure}

Since the sugar-phosphate backbone of the two probe DNAs are not
connected [see Fig.~\ref{fig:sequences}(a)], the DNA double
helix melts more like two duplexes. Therefore, a duplex formed
using two 12 base probes and a 16 base linker melts at a
temperature similar to that of a DNA double helix with eight
base pairs. The calculated thermodynamic values are given in
Table~\ref{table}. Figure~\ref{fig:Tm}(a) shows that the
experimentally observed free DNA melting temperatures are
described well by Eq.~(\ref{Tm}), with a deviation apparent as
the DNA duplex length approaches 12 bases; it is known that
Eq.~(\ref{Tm}) is accurate up to 12 bases and becomes less
accurate for longer sequences \cite{Sugimoto96}. In the
DNA-linked nanoparticles, however, $T_m$ does not increase
monotonically with linker DNA length. Specifically, the $T_m$ of
21 and 23 base linkers were found to be lower than that of 20
and 22 base linkers, respectively.

The $T_m$ and associated error bars for DNA linked to 10~nm
nanoparticles were determined from the averages and standard
deviations from four separate data sets. The error bars are
likely an overestimate of error in $T_m$ trend because
systematic error leads to a constant shift of $T_m$ with each
data set.  Parameters such as salt concentration, grafting
density, and linker concentration that will affect the $T_m$ are
kept constant within each data set by using the probe sample
from the same batch of solution and keeping the linker
concentration constant. To compare data within a given set,
where systematic errors are minimized, we plot the change in
melting temperature, $\Delta T_m = T_m(x) - T_m(x-1)$, versus
number of bases in the linker sequence, $x$.
Figure~\ref{fig:Tm}(b) demonstrates that linker lengths of 19,
21, and 23 are lower than expected, and this is true for both 10
and 20~nm nanoparticle systems. We found that the $T_m$
decreased when increasing linker length from from 20 to 21 and
from 22 to 23 bases. While an increase from 18 to 19 bases does
not result in a negative $\Delta T_m$, the $\Delta T_m$ is near
zero and significantly smaller than the predictions for free
DNA.

\begin{table}[b]
\vspace{-5ex}
\renewcommand\arraystretch{0.7}
\footnotesize
\begin{center}
\caption{Predicted thermodynamic parameters for unattached DNA duplexes}
\label{table}
\begin{ruledtabular}
\begin{tabular}{ccccccc}
  & & $\Delta H^0$ & $\Delta S^0$ & $\Delta G^0$ & [DNA] &  \\
  Linker\footnotemark[1] & Duplex\footnotemark[2]
  & \scriptsize ($\mbox{kcal} \over \mbox{mol}$)\footnotemark[3]
  & \scriptsize ($\mbox{cal} \over \mbox{mol} \cdot \mbox{K}$)\footnotemark[3]
  & \scriptsize ($\mbox{kcal} \over \mbox{mol}$)\footnotemark[4]
  & \scriptsize (10$^{\mbox{-}6}$M)\footnotemark[5]
  & $\overline{T}_m$\footnotemark[6] \\
  \hline
  16 & 8  & 60.05 & 162.0 & 11.75 & 7.83 & 23.9 \\
     & 8  & 62.25 & 169.1 & 11.83 &      &      \\
  18 & 9  & 68.05 & 183.9 & 13.22 & 6.64 & 29.6 \\
     & 9  & 70.25 & 191.0 & 13.30 &      &      \\
  19 & 9  & 68.05 & 183.9 & 13.22 & 6.61 & 32.1 \\
     & 10 & 78.25 & 212.9 & 14.77 &      &      \\
  20 & 10 & 76.05 & 205.8 & 14.69 & 6.65 & 34.6 \\
     & 10 & 78.25 & 212.9 & 14.77 &      &      \\
  21 & 10 & 76.05 & 205.8 & 14.69 & 6.73 & 36.8 \\
     & 11 & 86.25 & 234.8 & 16.24 &      &      \\
  22 & 11 & 84.05 & 227.7 & 16.16 & 6.57 & 38.8 \\
     & 11 & 86.25 & 234.8 & 16.24 &      &      \\
  23 & 11 & 84.05 & 227.7 & 16.16 & 6.40 & 40.5 \\
     & 12 & 94.25 & 256.7 & 17.71 &      &      \\
  24 & 12 & 92.05 & 249.6 & 17.63 & 6.37 & 42.3 \\
     & 12 & 94.25 & 256.7 & 17.71 &      &      \\
\end{tabular}
\end{ruledtabular}
\vspace{-3ex}
\normalsize
\footnotetext[1]{Each linker forms two duplexes. Linkers with an even
number of bases form equal length duplexes; odd numbered linkers
form duplexes differing in length by one base pair.}
\footnotetext[2]{The upper row of each linker indicates length of duplex formed
by starting at the 3$^\prime$ terminus and ending in the middle;
the lower row represents duplex starting from the middle to the
5$^\prime$ terminus.}
\footnotetext[3]{Calculated using experimentally determined parameters from~\cite{Sugimoto96}.}
\footnotetext[4]{Calculated using Eq.~(\ref{2ndlaw}).}
\footnotetext[5]{Determined using optical density of duplex DNA.}
\footnotetext[6]{$\overline{T}_m$ is the average of the two temperatures (from
the upper and lower rows) calculated using Eq.~(\ref{Tm}) with
0.3 M NaCl.}
\end{center}
\end{table}

This anomaly in $T_m$ is unique in the DNA-linked nanoparticle
system, and it does not result from the DNA duplex formation free energy,
as illustrated in Fig.~\ref{Tm}.  Careful examination of the linker
compositions reveals that, for linkers with an even number of
bases, the DNA duplexes formed are composed of a uniform DNA
length while those with an odd number of bases form duplexes
composed of two different DNA lengths.  For example, a 21 base
linker results in a connection composed of one 11 base and one 10
base duplex (11-10), whereas a 20 base linker results in a
(10-10). While an 11 base duplex is more stable than a 10 base
duplex, an assembly with the disorder introduced by the 11-10
duplex results in a lowering of the overall stability, hence the
melting temperature, of the system.
\begin{figure}[!t]
\begin{center}
\epsfig{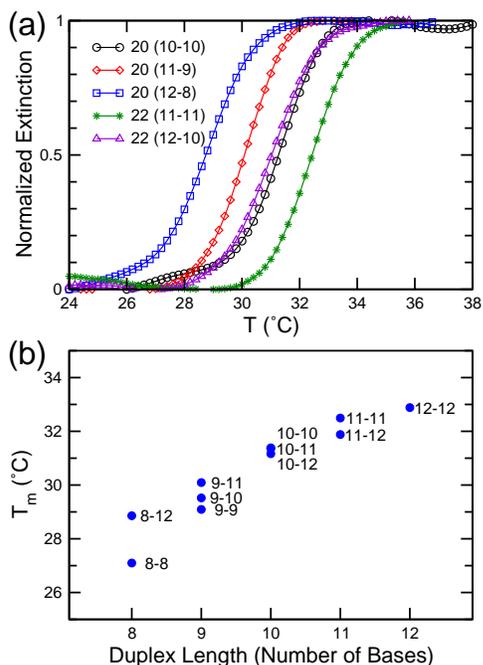}
\vspace{-0.3in}
\end{center}
\caption{(color online) (a) Melting curves at 260 nm for
aggregates formed using linkers with an equal number of bases,
but a different distribution of bases among the probe duplexes.
(b) Melting temperature as a function of the shortest DNA duplex
created using linkers of specific length and base distribution.}
\label{fig:symmetry}
\end{figure}

To verify that introducing a different DNA duplex length, which
introduces binding energy disorder, is responsible for lowering
the melting temperature, we tested linkers of the same length
with different sequences, as shown in
Fig.~\ref{fig:sequences}(b). We compared three 20 base linkers,
(10-10), (11-9), and (12-8). We found that
$T_m$(10-10)$>T_m$(11-9)$>T_m$(12-8). Typical melting curves for
these experiments are given in Fig.~\ref{fig:symmetry}(a).
Similar results are demonstrated for 22 base linkers,
$T_m$(11-11)$>T_m$(12-10). This leads to a conclusion that, in
DNA-linked nanoparticle assemblies, duplexes with different
lengths introduces binding energy disorder and, therefore, lower
the stability of the system. Since the energetics of
corresponding free DNA duplexes do not result in such an unusual
$T_m$ trend, we believe such an effect is largely a result of
entropic cooperativity \cite{Frenkel04a,Safran03a}.

The effect of having two DNA duplexes of different length on the
stability of DNA-linked nanoparticle assemblies is illustrated
in Fig.~\ref{fig:symmetry}(b). The stability is not simply
dominated by the shorter duplex, as demonstrated in
Fig.~\ref{fig:symmetry}(b), where $T_m$ is plotted as a function
of the shorter duplex length. In the case of a DNA linker
composed of one 10 base duplex,
$T_m$(12-10)$<T_m$(11-10)$<T_m$(10-10). This suggests that the
larger the length difference is between the two duplexes, the
less stable the aggregate will become. However, a decrease in
binding energy per base has a larger percentage effect on
shorter DNAs, so the unusual trend in $T_m$ for nine base and
shorter duplexes can been seen only quantitatively.

In summary, we found that disorder has a strong effect on the
stability in the DNA-linked nanoparticle assemblies. We have
demonstrated that using linker DNA that results in the presence of
two duplexes of different length and energy in such a system
lowers the overall stability of network formed. The interaction
energy is easily tunable by changing the DNA composition, which makes
this system particularly suitable for studying various aspects of
colloidal phase transitions.

We are grateful to Royce K. P. Zia for helpful discussions.
N.C.H. acknowledges the support from the Keck Center for
Computational and Structural Biology of the Gulf Coast
Consortia.

$^*$To whom correspondence should be addressed.

~Electronic address: chkiang@rice.edu
\vspace{-.1in}
\bibliography{disorder}

\end{document}